\documentclass{Interspeech}

\interspeechcameraready

\title{Bridging ASR and LLMs for Dysarthric Speech Recognition: Benchmarking Self-Supervised and Generative Approaches
}

\author[affiliation={1}]{Ahmed}{Aboeitta}
\author[affiliation={1}]{Ahmed}{Sharshar}
\author[affiliation={1}]{Youssef}{Nafea}
\author[affiliation={1,2}]{Shady}{Shehata}

\affiliation{NLP Department}{MBZUAI}{UAE}
\affiliation{}{Invertible AI}{UAE}
\email{\{ahmed.aboeitta,ahmed.sharshar,youssef.nafea,shady.shehata\}@mbzuai.ac.ae}


\usepackage{comment}
\usepackage{rotating}
\usepackage{tabularray}
\usepackage{booktabs}
\usepackage{multirow}
\usepackage[table,xcdraw]{xcolor}
\usepackage{graphicx}

\begin{document}

\maketitle

\begin{abstract}
    Dysarthric speech presents significant challenges for Automatic Speech Recognition (ASR) due to phoneme distortions and high variability. While self-supervised ASR models like Wav2Vec, HuBERT, and Whisper have shown promise, their effectiveness in dysarthric speech remains unclear. This study systematically benchmarks these models with different decoding strategies, including CTC, seq2seq, and LLM-enhanced decoding (BART, GPT-2, Vicuna). Our contributions include (1) benchmarking ASR architectures for dysarthric speech, (2) introducing LLM-based decoding to improve intelligibility, (3) analyzing generalization across datasets, and (4) providing insights into recognition errors across severity levels. Findings highlight that LLM-enhanced decoding improves dysarthric ASR by leveraging linguistic constraints for phoneme restoration and grammatical correction.
\end{abstract}

\section{Introduction}

Dysarthria is a motor speech disorder that disrupts articulation, pacing, and phoneme clarity, making automatic speech recognition (ASR) particularly challenging \cite{5,6, kim18e_interspeech}. Self-supervised ASR models such as HuBERT \cite{hsu2021hubert}, and Wav2Vec 2.0 \cite{baevski2020} have achieved strong performance on standard speech but struggle with dysarthric speech due to high phoneme variability. Traditional ASR approaches, whether CTC-based (e.g., Wav2Vec, HuBERT) or end-to-end models (e.g., Whisper \cite{radford2022robustspeechrecognitionlargescale}), face inherent limitations. CTC models misalign phonemes \cite{7}, while end-to-end models like Whisper lack linguistic constraints, causing grammatical errors. These challenges result in high Word Error Rates (WER) limiting ASR’s real-world usability for assistive technologies.

Prior research on dysarthric speech ASR has primarily focused on enhancing acoustic encoders through fine-tuning or domain adaptation, but decoding strategies remain underexplored \cite{chorowski2015}. While self-supervised ASR models, such as HuBERT, Wav2Vec, and Whisper, have been tested on dysarthric speech, they still exhibit high WER, particularly in moderate-to-severe cases \cite{wang2024enhancing, hu2024enhancing, shegal2023transfer, inproceedings}. While prior work has focused on improving feature representations, the role of decoding strategies in enhancing transcription remains underexplored.

Existing dysarthric ASR approaches employ either Connectionist Temporal Classification (CTC) decoding or end-to-end speech-to-text models, both of which have significant limitations \cite{Sawa2020An}. CTC-based models, such as Wav2Vec-CTC and HuBERT-CTC, assume phoneme independence, making them prone to misalignment errors when phonemes are distorted by dysarthria \cite{lee2025dypcldynamicphonemelevelcontrastive, Dingliwal2023Personalization}. Whisper's large-scale pretraining enhances robustness, but lacking linguistic constraints, it can produce syntactically or semantically incoherent transcriptions despite correct phoneme recognition \cite{Borgholt2020Do}.

Several hybrid ASR approaches have attempted to refine transcriptions using statistical language models, but these methods remain limited as they operate on ASR outputs rather than directly influencing the decoding process \cite{app12020903}. Since these approaches do not integrate linguistic constraints at the decoding stage, they are unable to fully mitigate phoneme distortions and grammatical inconsistencies.

Recent advances in ASR decoding have explored Large Language Models (LLMs) as integrated decoders, moving beyond traditional statistical models. Transformer-based architectures, such as sequence-to-sequence models, have demonstrated the potential to jointly model phonetic and linguistic constraints within ASR. Additionally, studies have investigated LLM-infused decoders, where models such as BART and GPT-3 are used as part of the ASR decoding pipeline, allowing for context-aware transcription generation \cite{ma2025asrerrorcorrectionusing, Higuchi2023Harnessing, Yu2023Connecting, Ling2023Adapting, Yu2023Connecting}.

This gap raises a critical question: Can LLM-Enhanced Decoding, rather than traditional CTC or seq2seq methods, improve dysarthric ASR by enforcing linguistic constraints and reducing phoneme-level errors? While past research has applied LLMs in ASR, their effectiveness as direct decoders for dysarthric speech remains unexamined. This study systematically investigates whether integrating LLMs within ASR decoding enhances transcription accuracy for dysarthric speakers.

To address this, we conduct a comprehensive benchmarking study evaluating the impact of LLM-Enhanced Decoding on dysarthric ASR. Our contributions are:

\begin{enumerate}
    \item Benchmarking ASR Architectures: We systematically compare HuBERT, Wav2Vec, and Whisper using different decoding strategies, including CTC-based, seq2seq, and LLM-Enhanced Decoding.

    \item Introducing LLM-based Decoding: We investigate the potential of BART, GPT-2, and Vicuna as integrated ASR decoders for dysarthric speech that enhance recognition accuracy and transcription intelligibility, rather than being used solely for post-ASR correction.

    \item Cross-Dataset Generalization Analysis: We evaluate model robustness by training on one dysarthric speech dataset (TORGO) and testing on another (UASpeech), assessing generalization capabilities.

    \item Comprehensive Discussion on Recognition Errors: We analyze WER trends across dysarthria severities (mild, moderate, severe) and discuss model performance variations under different ASR architectures.

\end{enumerate}

\section{Methodology}

\begin{figure*}[ht]
\centering
\includegraphics[width=\linewidth, height=8cm]{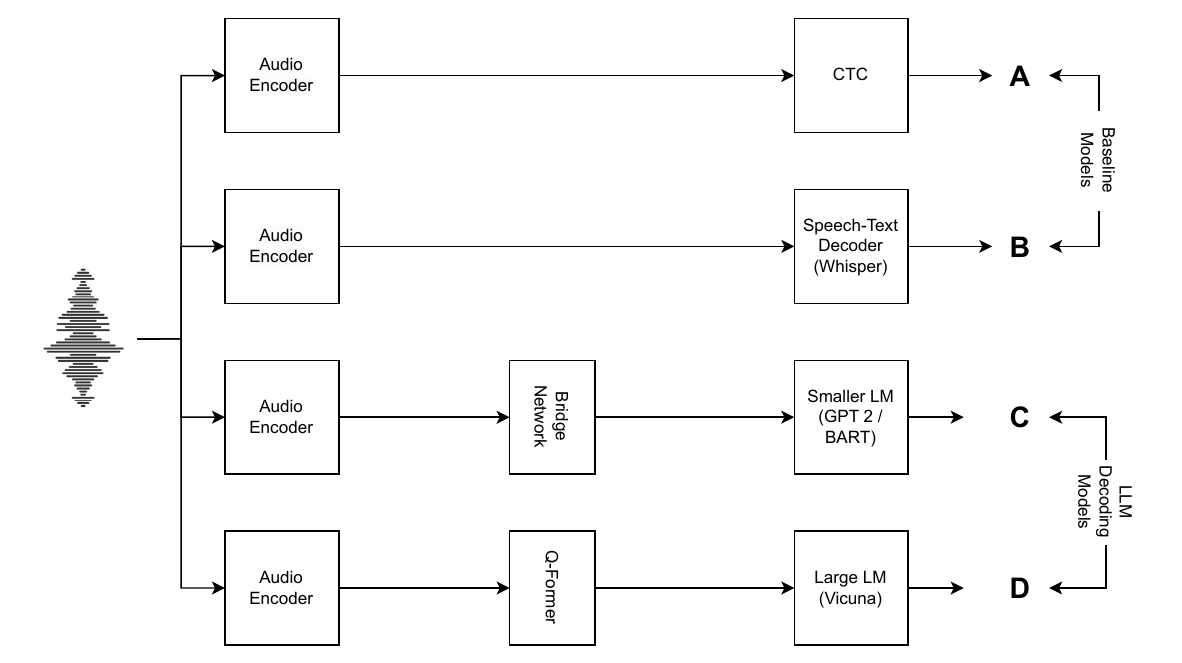}
\caption{Overview of the ASR architectures benchmarked in this study. (A) follows the traditional CTC-based decoding paradigm, where frame-wise phoneme prediction is performed directly from the encoder output. (B) employs Whisper, an end-to-end ASR model trained on large-scale speech-to-text data. (C) utilizes a Bridge Network, a neural network for adjusting feature dimensions, to connect smaller LLMs (GPT-2/BART) with the ASR encoder for refined transcription. (D) integrates a Large Language Model (Vicuna) via a Q-Former, allowing contextual and semantic-aware decoding, inspired by SALMONN \cite{tang2024salmonngenerichearingabilities}.}
\label{fig:approaches}
\end{figure*}

\subsection{Baseline ASR Models}
We evaluate two widely used self-supervised ASR architectures and their decoding mechanisms as baseline models for dysarthric speech recognition.

\textbf{(A) CTC-Based Decoding:} The traditional CTC-based approach directly maps acoustic features to text using frame-wise phoneme predictions. While efficient, CTC-based models lack the ability to enforce linguistic coherence, making them highly susceptible to errors in dysarthric speech recognition. This approach includes models like Wav2Vec 2.0 with CTC and HuBERT with CTC.

\textbf{(B) Whisper:} This end-to-end transformer-based ASR model is trained on large-scale speech-to-text data and implicitly learns both acoustic and linguistic structures during decoding. Whisper is used as a baseline model due to its state-of-the-art performance in various speech recognition tasks.

These ASR models excel in standard speech but lack linguistic modeling, leading to errors in dysarthric transcription. To address this, we introduce LLM-Enhanced Decoding models, as shown in Figure \ref{fig:approaches}.

\subsection{LLM-Enhanced Decoding Models}
To investigate the role of LLMs in dysarthric speech recognition, we integrate pretrained language models as decoders.This setup allows us to assess the impact of decoder strength on the accuracy of dysarthric speech recognition.

\textbf{(C) Small LLM-Based Decoding (GPT-2/BART with Bridge Network)}: In this approach, we used smaller language models such as GPT-2 and BART with a Bridge Network to align the ASR encoder’s output with the LLM’s text-based representations. The Bridge Network, implemented as a neural network, facilitates better representation transfer by modifying the size of the output audio features.

\textbf{(D) Large LLM-Based Decoding (Vicuna)}: In this approach, we integrate Whisper’s encoder with Vicuna, a conversational LLM, via a Q-Former. This configuration allows for semantic-aware decoding, leveraging Vicuna’s strong contextual reasoning capabilities to refine and correct dysarthric speech transcriptions dynamically. The Q-Former, inspired by the SALMONN framework \cite{tang2024salmonngenerichearingabilities}, tokenizes the output audio features before they are processed by Vicuna.

These models aim to determine whether LLM-Enhanced Decoding improves transcription intelligibility for dysarthric speakers by enforcing grammatical correctness and contextual understanding while addressing phoneme deletion and misalignment issues commonly found in dysarthric speech recognition. The impact of these decoding strategies is assessed in terms of transcription accuracy and error reduction across different levels of dysarthria severity.

\subsection{Datasets}
We use two benchmark dysarthric speech datasets: TORGO \cite{rudzicz2012torgo} and UASpeech \cite{kim2008dysarthric}. The TORGO dataset consists of 15 speakers (eight with dysarthria, seven typical), totaling 21 hours of English speech. Of this, 7.3 hours correspond to dysarthric speech and 13.7 hours to typical speech. The dataset includes short phrases and full sentences and is structured to ensure speaker independence: two-thirds of the dysarthric speakers' data are used for training, while the remaining one-third is reserved for testing. Typical speakers' data are excluded from training. TORGO categorizes speakers into three levels of intelligibility, allowing severity-based performance analysis.

UASpeech consists of 102.7 hours of speech from 29 speakers (16 dysarthric, 13 typical). Each speaker records 155 common words and 300 uncommon words, spread across three recording blocks. Following prior studies, we use Block 1 and Block 3 for training, while Block 2 serves as the test set, ensuring speaker independence. UASpeech provides four levels of intelligibility, enabling a finer-grained severity assessment.

\subsection{Experimental Setup}
We optimized the training process for each model through extensive hyperparameter tuning. Learning rates, warmup schedules, and the number of training epochs were carefully selected based on empirical performance across multiple configurations. Experiments were conducted on high-performance GPUs to handle the computational demands of LLM-Enhanced Decoding. TORGO was processed using an NVIDIA Quadro RTX 6000, while UASpeech required four NVIDIA A100 GPUs (40GB each) due to its larger size.

To systematically assess LLM-Enhanced Decoding, we evaluated baseline ASR models alongside LLM-augmented architectures. Baseline models included Wav2Vec 2.0 (large), HuBERT (large), and Whisper (large v2), representing both CTC-based and end-to-end ASR approaches. For LLM-Enhanced Decoding, we paired BART (large) and GPT-2 (large) with Wav2Vec 2.0 and HuBERT, while Vicuna (7.5B) was integrated with Whisper’s encoder. These models were selected to explore how autoregressive language modeling influences dysarthric ASR, with BART and GPT-2 providing text-based corrections and Vicuna enabling context-aware decoding.

All models were trained and evaluated on identical dataset partitions on dysarthric speech, ensuring that control (typical) speakers' data was not included in the training process to ensure fairness. Performance was assessed using WER across different dysarthria severity levels (mild, moderate, severe).

\section{Results \& Discussion}

\subsection{WER Comparison Across Models}

Self-supervised ASR models trained with CTC decoding (e.g., Wav2Vec-CTC, HuBERT-CTC) serve as baseline models for dysarthric speech recognition. These models rely on frame-wise phoneme classification without explicit linguistic modeling, making them highly sensitive to phoneme distortions and articulation inconsistencies. As shown in Table~\ref{tab:results_mix_uaspeech}, HuBERT-CTC (0.50 TORGO, 0.54 UASpeech) performs slightly better than Wav2Vec-CTC (0.53 TORGO, 0.54 UASpeech) due to its masked speech modeling, which enhances phoneme representations. However, both models exhibit high WER, confirming that CTC-based ASR struggles with phoneme misalignment and variability in dysarthric speech.

Whisper demonstrates a notable improvement over CTC-based models, reducing WER to 0.38 (TORGO) and 0.40 (UASpeech). This highlights the advantage of large-scale end-to-end training, which enables better generalization. However, despite this improvement, Whisper still shows performance degradation in moderate-to-severe dysarthria, indicating that acoustic modeling alone is insufficient for handling extreme phonetic distortions.

Integrating self-supervised speech encoders with LLM decoders further enhances performance. As observed in Table~\ref{tab:results_mix_uaspeech}, HuBERT-BART (0.30 TORGO, 0.32 UASpeech) and Wav2Vec-BART (0.32 TORGO, 0.35 UASpeech) show a substantial reduction in WER compared to both CTC-based and Whisper models. This highlights the effectiveness of linguistic modeling in decoding dysarthric speech.

Whisper-Vicuna achieves the lowest WER (0.21 TORGO, 0.26 UASpeech), leveraging context-aware error correction and semantic reconstruction enabled by LLM-Enhanced Decoding. Unlike Whisper, which relies solely on acoustic modeling, Vicuna refines transcriptions by incorporating linguistic structure, leading to improved transcription accuracy. Moreover, Whisper-Vicuna maintains the most consistent performance across datasets.

\begin{table}[ht]
\renewcommand{\arraystretch}{1.2}
\centering
\caption{WER results for different ASR architectures on TORGO and UASpeech.}
\label{tab:results_mix_uaspeech}
\begin{tabular}{@{}|ccc|@{}}
\hline
\multicolumn{1}{|c|}{\multirow{2}{*}{\textbf{Model}}} & \multicolumn{2}{c|}{\textbf{Dataset}}                   \\ \cline {2-3} 
\multicolumn{1}{|c|}{}                                & \multicolumn{1}{c|}{\textbf{Torgo}} & \textbf{UASpeech} \\ \hline
\multicolumn{3}{|l|}{\textbf{Baseline Models}}                                                                  \\ \hline
\multicolumn{1}{|c|}{Wav2Vec-CTC}                     & \multicolumn{1}{c|}{0.53}           & 0.54              \\
\multicolumn{1}{|c|}{Hubert-CTC}                      & \multicolumn{1}{c|}{0.50}            & 0.54              \\
\multicolumn{1}{|c|}{Whisper}                         & \multicolumn{1}{c|}{0.38}           & 0.40               \\ \hline
\multicolumn{3}{|l|}{\textbf{LLM-Enhanced Decoding Models}}                                              \\ \hline
\multicolumn{1}{|c|}{Wav2Vec-GPT}                     & \multicolumn{1}{c|}{0.59}           & 0.53              \\
\multicolumn{1}{|c|}{Hubert-GPT}                      & \multicolumn{1}{c|}{0.55}           & 0.50               \\
\multicolumn{1}{|c|}{Wav2Vec-BART}                    & \multicolumn{1}{c|}{0.32}           & 0.35              \\
\multicolumn{1}{|c|}{Hubert-BART}                     & \multicolumn{1}{c|}{0.30}            & 0.32              \\
\multicolumn{1}{|c|}{\textbf{Whisper-Vicuna}}         & \multicolumn{1}{c|}{\textbf{0.21}}  & \textbf{0.26}     \\ \hline
\end{tabular}
\end{table}

\subsection{Severity-Level Performance Analysis}

\begin{figure}[ht]
  \centering
  \includegraphics[width=0.5\textwidth]{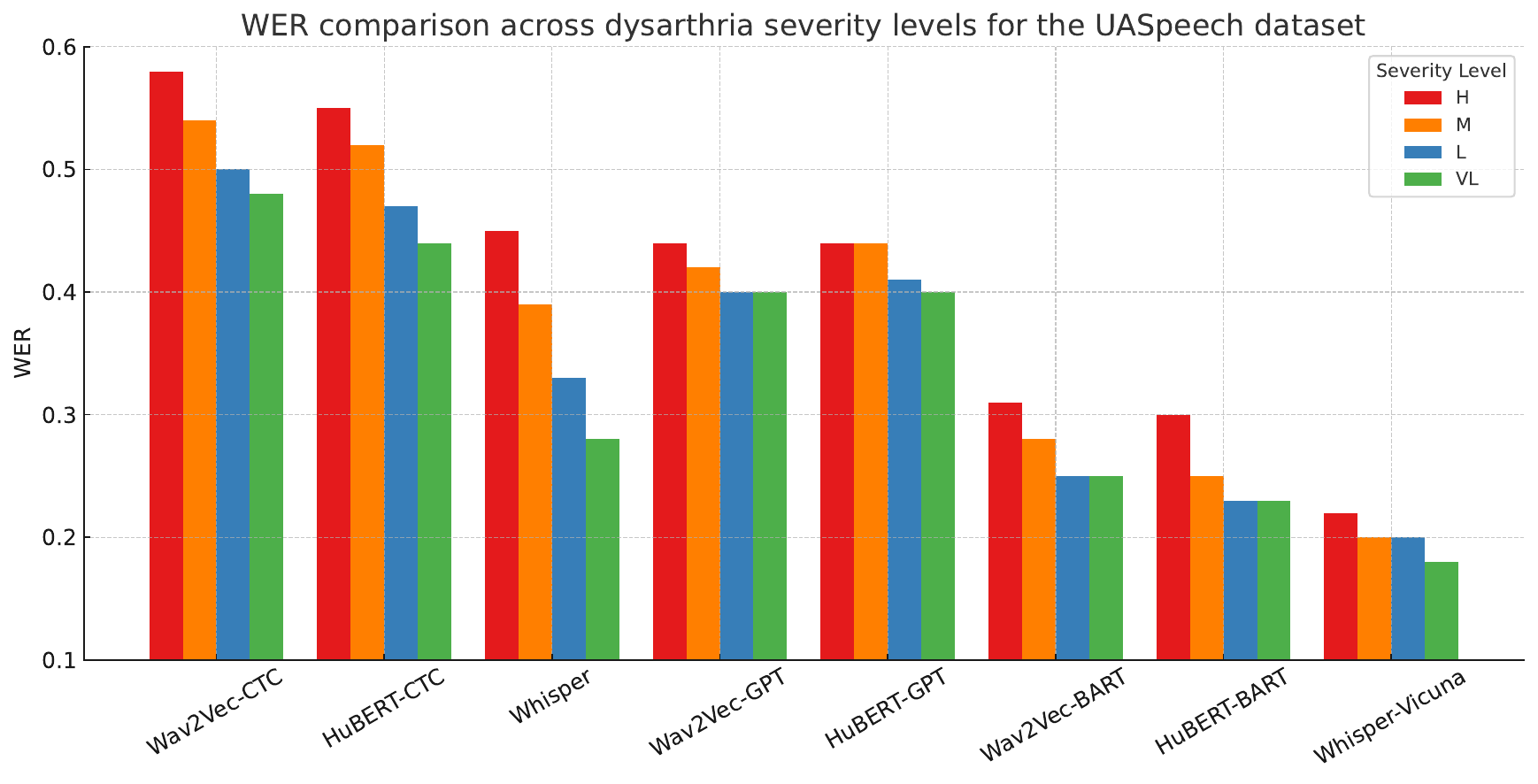}
  \caption{WER comparison across dysarthria severity levels for the UASpeech dataset}
  \label{fig:uaspeech_severity}
\end{figure}

\begin{figure}[ht]
  \centering
  \includegraphics[width=0.5\textwidth]{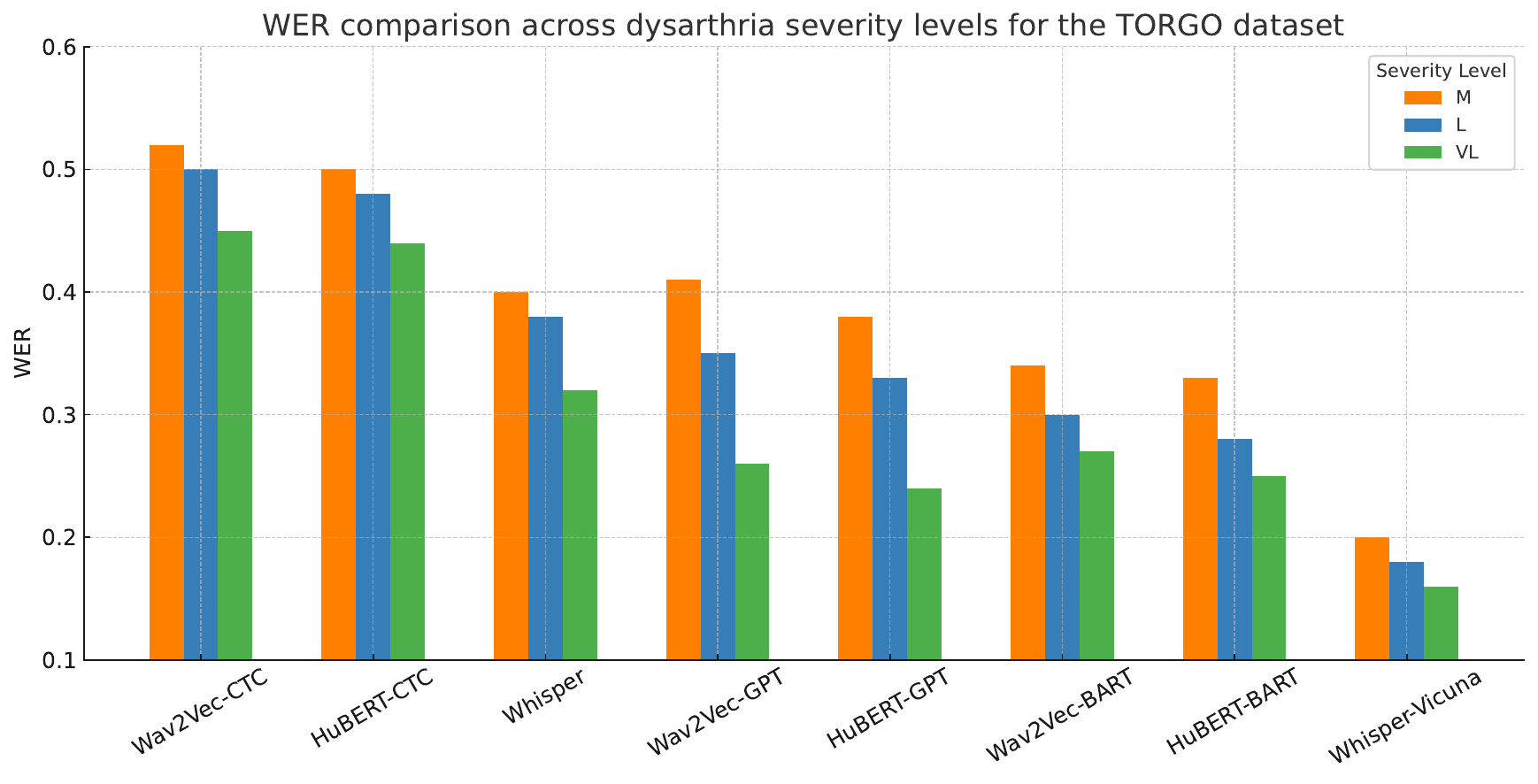}
  \caption{WER comparison across dysarthria severity levels for the TORGO dataset.}
  \label{fig:torgo_severity}
\end{figure}

Figures~\ref{fig:uaspeech_severity} and \ref{fig:torgo_severity} illustrate WER trends across Very Low (VL), Low (L), Moderate (M), and High (H) severity levels for all ASR models on TORGO and UASpeech. These results highlight each model’s robustness to increasing dysarthria severity.

CTC-based models (Wav2Vec-CTC, HuBERT-CTC) exhibit a steep WER increase with severity, reflecting their strong reliance on clear phoneme articulation. As distortions increase, their phoneme alignment weakens, leading to significantly higher errors.

Whisper achieves more stable WER due to end-to-end training, but its performance still deteriorates under severe dysarthria, showing that acoustic modeling alone is insufficient.

In contrast, LLM-decoder models (HuBERT-BART, Whisper-Vicuna) maintain lower WER across severity levels, highlighting the role of linguistic modeling in mitigating phoneme degradation. Whisper-Vicuna consistently achieves the lowest WER, leveraging context-aware error correction to compensate for phoneme-level distortions.

As shown in Figures~\ref{fig:uaspeech_severity} and \ref{fig:torgo_severity}, WER degradation rates differ by model. CTC-based models decline sharply, Whisper moderately, while LLM-based models, especially Whisper-Vicuna, remain the most robust.

\subsection{Error Analysis}

To assess transcription quality beyond WER, we analyze Character Error Rate (CER)  \begin{equation}
CER = \frac{S + D + I}{N}
\label{eq:cer}
\end{equation}
and sample transcriptions, offering a finer evaluation of phoneme distortions and semantic accuracy. We examine representative models: HuBERT-CTC (CTC-based), Whisper (end-to-end ASR), HuBERT-BART (LLM-enhanced), and Whisper-Vicuna, the best-performing model. Table~\ref{tab:cer_transcriptions} presents average CER across TORGO test sets, alongside transcriptions.

CTC-based models (e.g., HuBERT-CTC, CER = 0.28) suffer from frequent phoneme deletions and distortions, producing unintelligible transcriptions (e.g., "otl omner shrugg"). Whisper improves phoneme recognition (CER = 0.18) but often hallucinates words, leading to fluent yet incorrect transcriptions (e.g., "the hotel man").

LLM-enhanced models significantly reduce CER, demonstrating stronger phoneme restoration. HuBERT-BART (CER = 0.14) improves structure but retains minor errors (e.g., "the otel owner shrug"). Whisper-Vicuna (CER = 0.09) achieves the highest accuracy, closely matching the ground truth ("The hotel owner shrugged").

These results confirm that LLM-Enhanced Decoding mitigates phoneme distortions while improving linguistic coherence. While CER trends align with WER reductions, qualitative improvements in transcriptions highlight how linguistic modeling enhances intelligibility beyond numerical error rates.

\begin{table}[ht]
\renewcommand{\arraystretch}{1.2}
\centering
\caption{Character Error Rate (CER) and sample transcriptions from different ASR models on the TORGO dataset. The AVG CER is computed across all TORGO test sets, while the sample output demonstrates differences in transcription.}
\label{tab:cer_transcriptions}
\resizebox{0.48\textwidth}{!}{
\begin{tabular}{|cc|c|}
\hline
\multicolumn{1}{|c|}{\textbf{Model}}          & \textbf{AVG CER} & \textbf{Sample Output}            \\ \hline
\multicolumn{1}{|c|}{\textbf{HuBERT-CTC}}     & 0.28             & otl omner shrugg                  \\
\multicolumn{1}{|c|}{\textbf{Whisper}}        & 0.18             & the hotel man                     \\
\multicolumn{1}{|c|}{\textbf{HuBERT-BART}}    & 0.14             & The otel owner shrug              \\
\multicolumn{1}{|c|}{\textbf{Whisper-Vicuna}} & \textbf{0.09}    & The hotel owner shrugged          \\ \hline
\multicolumn{2}{|c|}{\textbf{Ground Truth}}                      & \textbf{The hotel owner shrugged} \\ \hline
\end{tabular}
}
\end{table}

\subsection{Generalization Across Datasets}

To evaluate cross-dataset generalization, models trained on one dataset were tested on the other. As shown in Table~\ref{tab:generalization}.

CTC-based models, particularly HuBERT-CTC, show the highest degradation, with WER reaching 1.86 when trained on UASpeech and tested on TORGO. This highlights their sensitivity to domain shifts. Whisper-Vicuna achieves the lowest WER in both setups but still degrades notably, underscoring the challenge of adapting to unseen dysarthria variations.

These results reveal phonetic and acoustic variability across datasets, limiting model robustness beyond training conditions. While LLM-Enhanced Decoding improves performance, generalization remains a key challenge, reinforcing the need for more diverse datasets and adaptive learning strategies.

\begin{table}[ht]
\renewcommand{\arraystretch}{1.2}
\centering
\caption{Cross-dataset generalization results: WER when models trained on one dataset were tested on another.}
\label{tab:generalization}
\resizebox{0.48\textwidth}{!}{ 
\begin{tabular}{|c|c|c|}
\hline
\textbf{Model}          & \textbf{\begin{tabular}[c]{@{}c@{}}Trained on Torgo\\ Tested on UASpeech\end{tabular}} & \textbf{\begin{tabular}[c]{@{}c@{}}Trained on UASpeech\\ Tested on Torgo\end{tabular}} \\ \hline
\textbf{Hubert-CTC}     & 1.56                                                                                   & 1.86                                                                                   \\
\textbf{Whisper}        & 1.20                                                                                    & 1.10                                                                                    \\
\textbf{Hubert-BART}    & 0.98                                                                                   & 0.99                                                                                   \\
\textbf{Whisper-Vicuna} & \textbf{0.87}                                                                                   & \textbf{0.88}                                                                                   \\ \hline
\end{tabular}
}
\end{table}

\section{Limitations}
While LLM-enhanced decoding improves dysarthric ASR, challenges remain. Cross-dataset generalization is poor, with models showing high WER increases on unseen data. Even LLM-assisted models, like Whisper-Vicuna, though more robust, still degrade across datasets. Limited dysarthric speech data further hinders robustness, as small datasets impact generalization compared to large ASR corpora. Data augmentation or synthetic speech generation could help. Lastly, architectural constraints restrict broader encoder-decoder evaluations, such as Whisper Encoder + BART.

\section{Conclusion}
This study benchmarks self-supervised ASR models (Wav2Vec, HuBERT, Whisper) with LLM-enhanced decoding (BART, GPT-2, Vicuna) for dysarthric speech recognition. Unlike prior work focusing only on encoders, it integrates LLMs in the decoding stage to improve transcription intelligibility. Our contributions include (1) benchmarking ASR architectures, (2) introducing LLM-based decoding, (3) analyzing cross-dataset generalization, and (4) studying recognition errors across severity levels. Results show that CTC-based models struggle due to phoneme distortions, while Whisper performs better but lacks strong linguistic modeling. LLM-assisted models, particularly Whisper-Vicuna, significantly reduce WER by leveraging linguistic context for better decoding, supporting our hypothesis that LLM-enhanced decoding improves phoneme restoration and grammatical accuracy.

Future work aims to expand dysarthric speech datasets and incorporate multimodal approaches to enhance recognition. A unified ASR-LLM framework should be designed to accommodate diverse encoder-decoder configurations.

\bibliographystyle{IEEEtran}
\bibliography{mybib}

\end{document}